\documentclass[%
reprint,
superscriptaddress,
 amsmath,amssymb,
 aps,
floatfix,
]{revtex4-2}

\usepackage[dvipsnames,svgnames,x11names,hyperref]{xcolor}
\usepackage{siunitx}

\usepackage{graphicx}
\usepackage[centering,hmargin=25mm,tmargin=28mm,bmargin=28mm]{geometry}
\usepackage{dcolumn}
\usepackage{bm}
\usepackage{hyperref}
\usepackage[mathlines]{lineno}
\usepackage{mathrsfs}
\usepackage[normalem]{ulem}

\usepackage[normalem]{ulem}
\hypersetup{colorlinks=true, 
	linkcolor={blue!75!black!80!yellow},
	citecolor={blue!75!black!80!yellow}, 
	urlcolor=black 
	}
\makeatletter \renewcommand\@make@capt@title[2]{%
\@ifx@empty\float@link{\@firstofone}{\expandafter\href\expandafter{\float@link}}%
\sffamily{\textbf{#1}}\@caption@fignum@sep#2 }

\begin{document}
\preprint{APS/123-QED}

\title{Spin emitters beyond the point dipole approximation in nanomagnonic cavities}

\author{Derek S. Wang}
\email{derekwang@g.harvard.edu}
\affiliation{Harvard John A. Paulson School of Engineering and Applied Sciences, Harvard University, Cambridge, MA 02138, USA}

\author{Tom\'{a}\v{s} Neuman}
\affiliation{Harvard John A. Paulson School of Engineering and Applied Sciences, Harvard University, Cambridge, MA 02138, USA}

\author{Prineha Narang}
\email{prineha@seas.harvard.edu}
\affiliation{Harvard John A. Paulson School of Engineering and Applied Sciences, Harvard University, Cambridge, MA 02138, USA}

\begin{abstract}
Control over transition rates between spin states of emitters is crucial in a wide variety of fields ranging from quantum information science to the nanochemistry of free radicals. We present an approach to drive a both electric and magnetic dipole-forbidden transition of a spin emitter by placing it in a nanomagnonic cavity, requiring a description of both the spin emitter beyond the point dipole approximation and the vacuum magnetic fields of the nanomagnonic cavity with a large spatial gradient over the volume of the spin emitter. We specifically study the SiV$^-$ defect in diamond, whose Zeeman-split ground states comprise a logical qubit for solid-state quantum information processing, coupled to a magnetic nanoparticle serving as a model nanomagnonic cavity capable of concentrating microwave magnetic fields into deeply subwavelength volumes. Through first principles modeling of the SiV$^-$ spin orbitals, we calculate the spin transition densities of magnetic dipole-allowed and -forbidden transitions and calculate their coupling rates to various multipolar modes of the nanomagnonic cavity. We envision using such a framework for quantum state transduction and state preparation of spin qubits at GHz frequency scales.
\end{abstract}
\date{\today}

\maketitle

\section{Introduction}
Precise control over transition rates between states of an emitter and, thus, absorption and emission of phonons, photons, and magnons
is crucial for a variety of applications ranging from quantum information processing to energy harvesting in artificial and biological structures \cite{Quidant2010, Song2012, Tame2013, Hayat2011, Rivera2016}. These transition rates are governed by selection rules that generally assume that these external fields are plane waves with negligible spatial gradient and that the emitter can be represented as a point dipole \cite{Cohen2009, Yang2011, Lee2011}. Violating either or both of these approximations, however, can lead to exotic optical and chemical phenomena, such as multipolar transitions that are faster than dipolar ones, multiquanta emission, efficient spin-flip processes, and spatially-dependent near-field emission \cite{Andersen2011, Takase2013, cotrufo2015, Rivera2016, Neuman2018pointdipole}. Further experimental development of such processes to tailor transition rates may lead to facile generation of highly entangled multipartite states, for instance \cite{Huber2018, Zeeshan2019, Wang2020}.

Although previous research in breaking selection rules has been largely focused on the interaction between plasmonic systems with emitters in the optical frequency range, there have been comparatively fewer studies on breaking selection rules in spin emitters \cite{Imamoglu2003, Cerletti2005, Lehmann2007, Trauzettel2007, Muller2008, Kloeffel2013, sukachev2017sivspinqubit, Pingault2017, becker2018alloptical} with transition frequencies in the microwave range, where magnetic dipolar transitions are inherently orders of magnitude slower than their electric dipolar counterparts \cite{Cohen-Tannoudji:101367}. Spin emitters are of large technological interest given, for instance, the proposed usage of defect spin emitters as spin-photon interfaces in quantum information processing systems \cite{Kurizki2015, Li2015, Dreau2018, Awschalom2018-en, christle2015isolated} and radical pair-based mechanisms as the source of biological magneto-reception \cite{Grissom1995, Rodgers2009}. 

\begin{figure}[tbhp]
\centering
\includegraphics[width=0.98\linewidth]{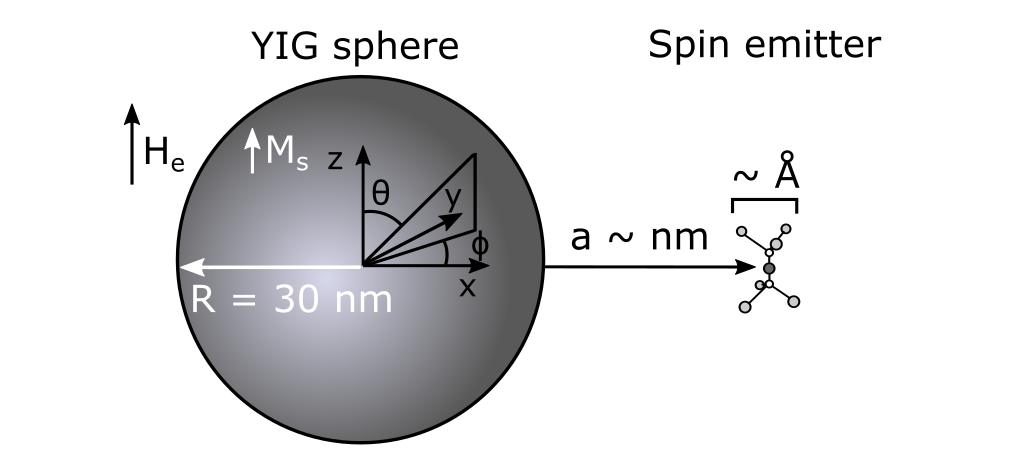}
\caption{The spatially-varying vacuum magnetic field of a nanomagnonic cavity realized with a magnetic nanoparticle with radius $R$ drives the spin- and strain-forbidden transition  between the $|e^+,\uparrow\rangle$ and $|e^-,\downarrow\rangle$ states of a model spin emitter, the SiV$^-$ defect center in diamond, placed $a$ away from the surface of the nanoparticle.
}
\label{fig:schematic}
\end{figure}

In this \textit{Article}, we present an approach to couple isolated spin emitters to fields of magnon modes with length scales on the order of single molecular or defect emitters. The magnon modes are realized by ferromagnetic or ferrimagnetic nanoparticles as nanomagnonic cavities that concentrate microwave magnetic fields into deeply subwavelength volumes. This configuration is illustrated in Fig. \ref{fig:schematic}, where the spin emitter is positioned close to the nanoparticle surface and efficiently coupled via fields of dipolar or higher order magnon modes. The large spatial gradients of coupled magnon modes requires description of the emitters beyond the point-dipole approximation and enables selection rule-breaking of orbital-spin transitions, This phenomenon is analogous to those realized in studies that have demonstrated breaking of electric dipole-based selection rules in plasmonic systems \cite{Andersen2011, Takase2013, cotrufo2015, Rivera2016, Neuman2018pointdipole}. While Ref. \cite{Cohen2009, Yang2011} consider how magnetic nanostructures can generate magnetic fields with large gradients and catalyze intersystem crossing in molecular radical pairs, neither the full spectral profile of the magnetic nanostructure nor the spatial variation of the spin emitters was considered. To demonstrate selection rule breaking, we show a non-zero transition rate between the logical $|0\rangle$ and $|1\rangle$ states in the ground state manifold of the negative silicon vacancy (SiV$^-$) in diamond, a leading materials candidate for spin-photon coupling in solid-state qubits for quantum technologies \cite{sukachev2017sivspinqubit, Pingault2017, becker2018alloptical, harris2019group, ciccarino2020strong, Neuman2020}, coupled to a spherical nanoparticle of yttrium iron garnet (YIG) that serves as the nanomagnonic cavity. This transition is typically forbidden under electric and magnetic dipole selection rules. We show that this magnon-spin coupling rate can reach kHz frequency scales with the potential to enter the strong magnon-spin coupling regime upon the development of lower-loss magnetic materials. We envision leveraging these phenomena to more flexibly manipulate the quantum states of spin qubits or to mediate qubit-qubit interactions necessary for quantum technologies. 

\section{SiV$^-$ as a model spin emitter}

\begin{figure}[tbhp]
\centering
\includegraphics[width=0.98\linewidth]{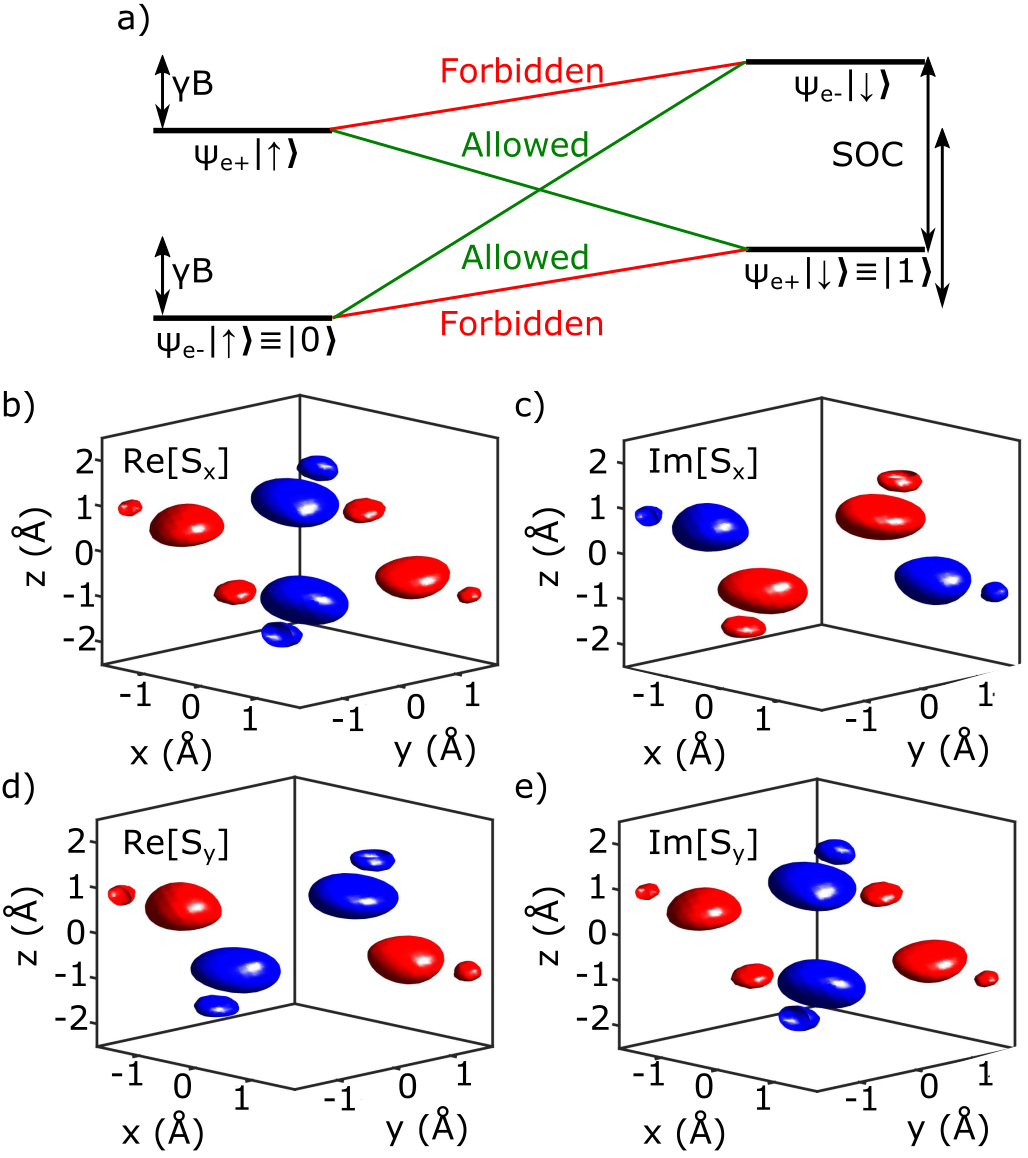}
\caption{\textbf{(a)} Energy-level diagram of the ground state of the SiV$^-$ defect center in diamond. The magnetic dipole-allowed and forbidden transitions are labelled in green and red, respectively. The real and imaginary parts of the components of the spin transition densities \textbf{(b)}-\textbf{(c)} $S_x(\mathbf{r})$ and \textbf{(d)}-\textbf{(e)} $S_y(\mathbf{r})$. $S_z(\mathbf{r})$ is 0.
}
\label{fig:emitter}
\end{figure}

As an example of a spin emitter with a transition forbidden by magnetic diple-based selection rules, we study the SiV$^-$ defect in diamond, a leading spin-photon interface candidate for various quantum information processing technologies \cite{sukachev2017sivspinqubit, Pingault2017, becker2018alloptical}. The SiV$^-$ is a split-vacancy defect that consists of the silicon atom located between two adjacent vacant sites in diamond, resulting in $D_{3d}$ symmetry \cite{Hepp2014, Thiering2018}. The degeneracy of the ground state manifold is broken by spin-orbit coupling that has been experimentally observed at $\sim$50 GHz. This value can be closely reproduced with \emph{ab initio} density functional theory calculations with a correction due to the Jahn-Teller effect, although this effect is not necessary to predict the orbital and spin character \cite{Thiering2018}. The resulting ground state manifold---with lower energy degenerate states $\psi_{e^+}(\mathbf{r})|\downarrow\rangle$ and $\psi_{e^-}(\mathbf{r})|\uparrow\rangle$ and higher energy degenerate $\psi_{e^+}(\mathbf{r})|\uparrow\rangle$ and $\psi_{e^-}(\mathbf{r})|\downarrow\rangle$---can be coupled together via microwave drive. The degeneracies of the lower and upper levels of the ground state manifold can be broken in the presence of longitudinal magnetic field along the [111] direction. The lowest two states $\psi_{e^+}(\mathbf{r})|\downarrow\rangle$ and $\psi_{e^-}(\mathbf{r})|\uparrow\rangle$ are of special interest, as they comprise a spin qubit $|0\rangle$ and $|1\rangle$, respectively, used successfully in Refs. \cite{bhaskar2019experimental, nguyen2019nuclearoptics}. Notably, transitions within this spin qubit are forbidden via homogeneous magnetic field due to the opposite parities of the spatial orbitals $\psi_{e^+}(\mathbf{r})$ and $\psi_{e^+}(\mathbf{r})$ and via a homogeneous electric field due to their opposite spins \cite{Neuman2020phononicbus}. This transition, however, can be allowed with a combination of transverse magnetic field, such that the eigenstates are mixtures of $\psi_{e^+}(\mathbf{r})|\downarrow\rangle$ and $\psi_{e^-}(\mathbf{r})|\uparrow\rangle$, and/or strain \cite{udvarheliy2018spinstrain, Meesala2018, Neuman2020phononicbus, Maity}. We show here that this transition can be driven solely with the vacuum magnetic field with a large spatial gradient of a nanomagnonic cavity, physically realized here as a ferrimagnetic YIG nanoparticle.

\section{Magnonic modes of a magnetic nanosphere}
\begin{figure}[tbhp]
\centering
\includegraphics[width=0.98\linewidth]{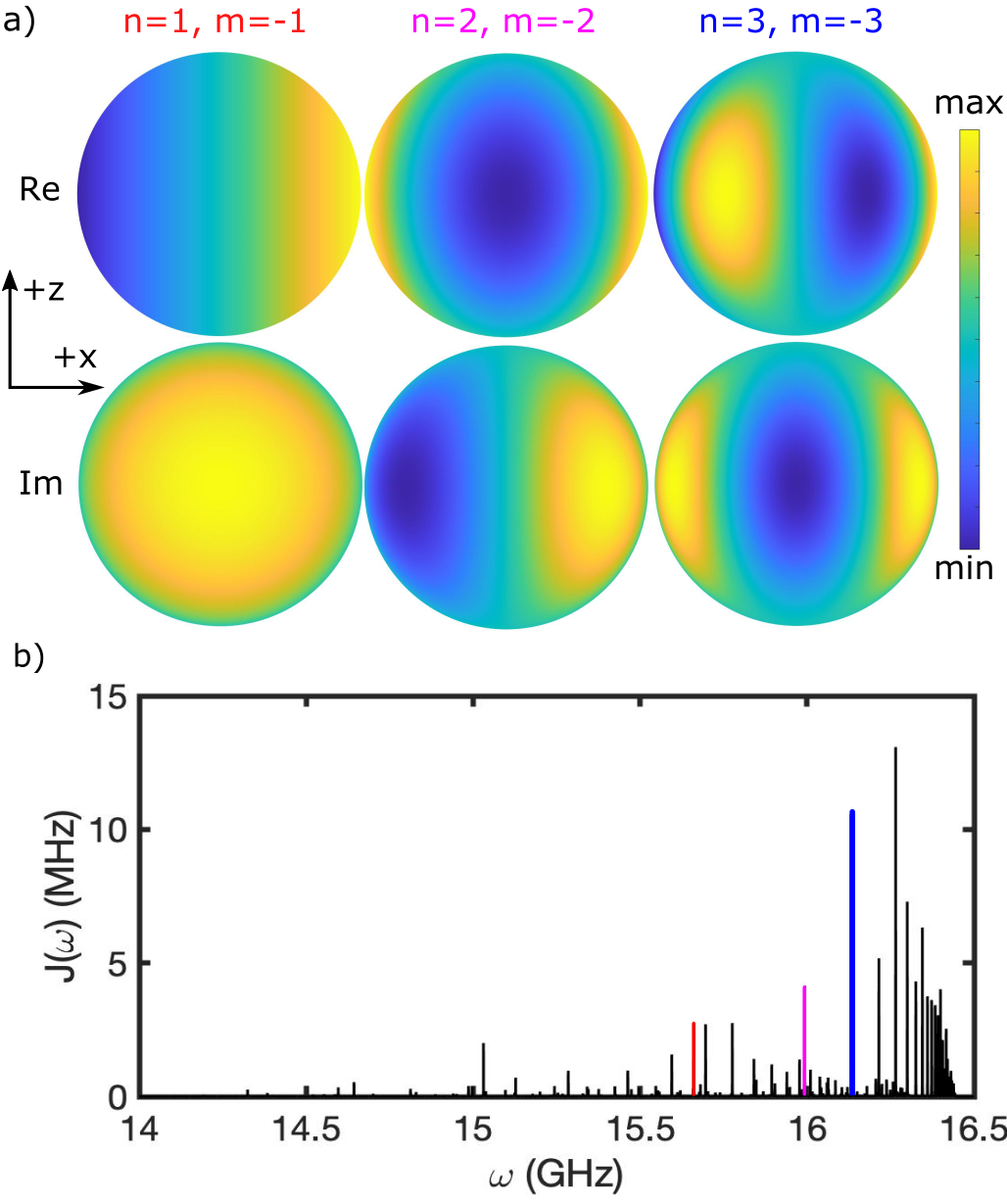}
\caption{\textbf{(a)} Real and imaginary components of the surface potential $\phi(R)$ looking along the $y$-axis of the equatorial multipolar modes of the nanomagnonic cavity, specifically the dipolar mode where $n=1$ and $m=-1$, quadrupolar mode where $n=2$ and $m=-2$, and octupolar mode where $n=3$ and $m=-3$. \textbf{(b)} Spectral density $J(\omega)$ of the nanomagnonic cavity coupled to a magnetic point dipole including multipolar terms up to $n=25$ for $R=30$ nm, $a=5$ nm. The relevant dipolar, quadrupolar, and octupolar modes are highlighted in red, pink, and blue, respectively.
}
\label{fig:cavity}
\end{figure}

We briefly present the magnonic modes of a homogeneously magnetized nanosphere with an external magnetic field aligned along the magnetization direction $z$ following the treatment of Walker \cite{walker1957magnetostatic, walker1958resonant} and Fletcher \cite{Fletcher1959ferrimagnetic} and discussed more extensively in Ref. \cite{Neuman2020Nanomagnonics}. In particular we seek the solutions for a magnon mode:
\begin{align}
    (1+\chi)\left(\frac{\partial^2 \phi}{\partial x^2}+\frac{\partial^2 \phi}{\partial y^2}\right)+\frac{\partial^2 \phi}{\partial z^2}=0,
\end{align}
where the magnetic scalar potential $\phi$ is linked with the quasi-static magnetic field by ${\bf H}(\omega)=-\nabla\phi(\omega)$ and $\omega$ is the magnon frequency. $\chi\equiv \chi_{xx}=\chi_{yy}$ and $\mathrm{i}\kappa\equiv \chi_{xy}=\chi_{yx}^*$ are the frequency-dependent components of the susceptibility tensor $\bm{\chi}(\omega)$ defined by the magnetic response to external fields in the Landau-Lifshitz-Gilbert equation \cite{Neuman2020Nanomagnonics}. The inputs required to calculate the susceptibility tensor $\bm{\chi}(\omega)$ are the gyromagnetic ratio $\gamma\approx 176$\,\si{\giga\radian\cdot \tesla^{-1}}, the phenomenological damping parameter $\Gamma=1$ Mrad/s, external magnetic field $H_e=0.5$ T, and saturation magnetization $\mu_0 M_{\rm s}=0.178$\,\si{\tesla} \cite{Neuman2020Nanomagnonics}, which closely correspond to the values observed experimentally in YIG \cite{Wu2010yig}. 

For the outer region of the sphere we define the field in the standard spherical coordinates as shown in Figure \ref{fig:schematic}:
\begin{align}
    x&=r\sin(\theta)\cos(\varphi),\\
    y&=r\sin(\theta)\sin(\varphi),\\
    z&=r\cos(\theta),
\end{align}
where the magnetic scalar potential outside the sphere $\phi_\mathrm{out}$ is
\begin{align}
    \phi_{\rm out}=B_{nm}r^{-n-1}P_n^m(\cos\theta)e^{im\varphi},
\end{align}
where $P_n^m(\cos\theta)$ is the associated Legendre polynomial and each magnon mode is defined by a pair of indices $n$ and $m$. The magnetic scalar potential inside the sphere $\phi_\mathrm{in}$ can be expressed in ellipsoidal coordinates:
\begin{align}
    x&=R\sqrt{-\chi}\sqrt{1-\xi^2}\sin(\eta)\cos(\varphi),\\
    y&=R\sqrt{-\chi}\sqrt{1-\xi^2}\sin(\eta)\sin(\varphi),\\
    z&=R \sqrt{\frac{\chi}{1+\chi}}\xi \cos(\eta),
\end{align}
resulting in the sought solution:
\begin{align}
    \phi_{\rm in}=A_{nm}P_n^m(\xi)P_n^m(\cos\eta)e^{i m\varphi}.
\end{align}
The solutions must fulfill the boundary conditions \cite{Fletcher1959ferrimagnetic}:
\begin{align}
    \left( \phi_{\rm in} \right)_R=\left( \phi_{\rm out} \right)_R, 
\end{align}
and 
\begin{align}
    \left( \frac{\partial \phi_{\rm out}}{\partial r} \right)_R&=(1+\chi\sin^2\theta)\left( \frac{\partial \phi_{\rm in}}{\partial r} \right)_R\nonumber\\
    &+\frac{\chi\sin\theta\cos\theta}{R}\left( \frac{\partial \phi_{\rm in}}{\partial \theta} \right)_R\nonumber\\
    &+\frac{i\kappa}{R}\left( \frac{\partial \phi_{\rm in}}{\partial \varphi} \right)_R. 
\end{align}
Note that the sign convention $e^{-i\omega t}$ has been used.
After inserting $\phi_{\rm in}$ and $\phi_{\rm out}$ into the boundary conditions and noticing that on the particle surface $\xi_0=[(1+\chi)/\chi]^{1/2}$ and $\eta=\theta$, we obtain:
\begin{align}
    B_{mn}=-A_{nm}\left[ \frac{\xi_0P_n^m{}'(\xi_0)-m\kappa P_n^m(\xi_0)}{n+1} \right]R^{n+1}\label{eq:se1}
\end{align}
and 
\begin{align}
    B_{nm}=A_{nm}R^{n+1}P_n^m(\xi_0).\label{eq:se2}
\end{align}
Equations\,\eqref{eq:se1} and \eqref{eq:se2} must be fulfilled simultaneously, therefore posing a condition on the acceptable value of frequency $\omega$ appearing in $\xi_0(\omega)$ [i.e. $\chi(\omega)$] and $\kappa(\omega)$. The acceptable frequencies (resonance frequencies) follow from the secular equation:
\begin{align}
    (n+1-m\kappa)P_n^m(\xi_0)+\xi_0P_n^{m}{}'(\xi_0)=0.\label{eq:sec}
\end{align}
The secular equation [Eq.\,\eqref{eq:sec}] can be solved for low values of $n$ and $m$ analytically as shown in Ref. \cite{Fletcher1959ferrimagnetic}. The magnonic modes to which we couple the spin emitter are plotted in Fig. \ref{fig:cavity}(a), where $n=1$, 2, and 3 and $m=-n$ corresponding to the equatorial dipolar, quadrupolar, and octupolar modes, respectively.

\section{Magnon-emitter coupling}
To determine the magnon-emitter coupling rate $g$, we first obtain the spin transition density ${\mathbf S}({\bm r})$ for the forbidden transition, where the $\alpha \in \{x,y,z\}$ component can be expressed as
\begin{align}
   {S}_\alpha(\mathbf{r})=\psi_{e^+}(\mathbf{r})^* \psi_{e^-}(\mathbf{r}) \langle \downarrow | \sigma_\alpha | \uparrow \rangle,
\end{align}
where the Pauli vector ${\mathbf \sigma}=(\sigma_x, \sigma_y, \sigma_z)$. We plot each component of the spin transition densities in Fig. \ref{fig:emitter}, where the spin orbitals are calculated as described in Appendix \ref{app:orbitalextraction}.

The spectral density $J(\omega)$ for a spatially-delocalized spin emitter can then be written as
\begin{align} 
    J(\omega)&=\frac{\mu_0\mu_{\rm B}^2}{\hbar\pi}\nonumber\\
    &{\rm Im}\left\{  \iint {\bf S}^\ast({\bf r}')\cdot k_0^2 {\bf G}_{\rm m}({\bf r}', {\bf r}, \omega) \cdot {\bf S}({\bf r}){\rm d}^3{\bf r}{\rm d}^3{\bf r}'\right\},\label{eqs:specdenJgen}
\end{align}
where $\mu_{\rm B}$ is the Bohr magneton and ${\bf G}_{\rm m}({\bf r}', {\bf r}, \omega)$ is the magnetostatic Green's tensor. This expression is generalized from the expression for the coupling between a quantized magnon mode and a point-like spin emitter \cite{Neuman2020Nanomagnonics}. To evaluate the Green's tensor ${\bf G}_{\rm m}({\bf r}', {\bf r}, \omega)$ we solve the Poisson's equation in spherical geometry as described in Ref. \cite{Neuman2020Nanomagnonics} using the expansion of the magnetic field in the magnon modes. The spectral function $J(\omega)$ determines the dynamics of the spin excitation, with $c_{\rm e}$ being the excited-state amplitude, as
\begin{align}
    \dot{\tilde{c}}_{\rm e} = -\int_0^tf(t-t'){\tilde{c}}_{\rm e}(t')\,{\rm d}t',\label{eqs:IDEdef}
\end{align}
where ${\tilde{c}}_{\rm e}(t')e^{-i\omega_0 t'}={{c}}_{\rm e}(t')$ and
\begin{align}
    f(t-t')=\int_{-\infty} ^\infty J(\omega) e^{i(\omega_0-\omega)(t-t')}{\rm d}\omega. \label{eq:ftt}
\end{align}

Assuming that the spectral density $J(\omega)$ within $\Gamma$ around the frequency $\omega_{n,m}$ of the magnon mode defined by multipole index $n$ and angular momentum index $m$ to which the spin transition frequency is tuned corresponds only to this single magnon mode, valid in the limit of small damping $\Gamma=1$ Mrad/s for YIG, we can determine the coupling rate $g_{n,m}$ by fitting
\begin{align}
    \tilde{J}(\omega)=\frac{1}{2\pi}\frac{\Gamma g_{n,m}^2}{(\Gamma/2)^2+(\omega-\omega_{n,m})^2}
\end{align}
to the full magnetostatic solution for $J(\omega)$.
To calculate the coupling rate for the magnetic dipole-allowed transition, we can repeat this procedure after first writing the spin transition density as
\begin{align}
   {S}_\alpha(\mathbf{r})=\psi_{e^+}(\mathbf{r})^* \psi_{e^+}(\mathbf{r}) \langle \downarrow | \sigma_\alpha | \uparrow \rangle. \label{eqs:allowedSpinDensity}
\end{align}
In Fig. \ref{fig:cavity}(b), for a nanosphere with radius $R=30$ nm and magnetic point dipole position $a=5$ nm away, we plot the spectral density $J(\omega)$ under the dipolar approximation, where $\psi_{e^+}(\mathbf{r})^* \psi_{e^+}(\mathbf{r})$ in Eq. \eqref{eqs:allowedSpinDensity} can be approximated by a delta function and Eq. \eqref{eqs:specdenJgen} reduces to the form in Ref. \cite{Neuman2020Nanomagnonics}. Note that the three modes plotted in Fig. \ref{fig:cavity}(a) have relatively large values of $J(\omega)$ and their peaks are are well-separated from other magnon frequencies within their linewidths $\sim \Gamma$, in contrast to the near-continuum of higher-order magnon modes above $\sim$16.3 GHz.

\begin{figure}[tbhp]
\centering
\includegraphics[width=0.98\linewidth]{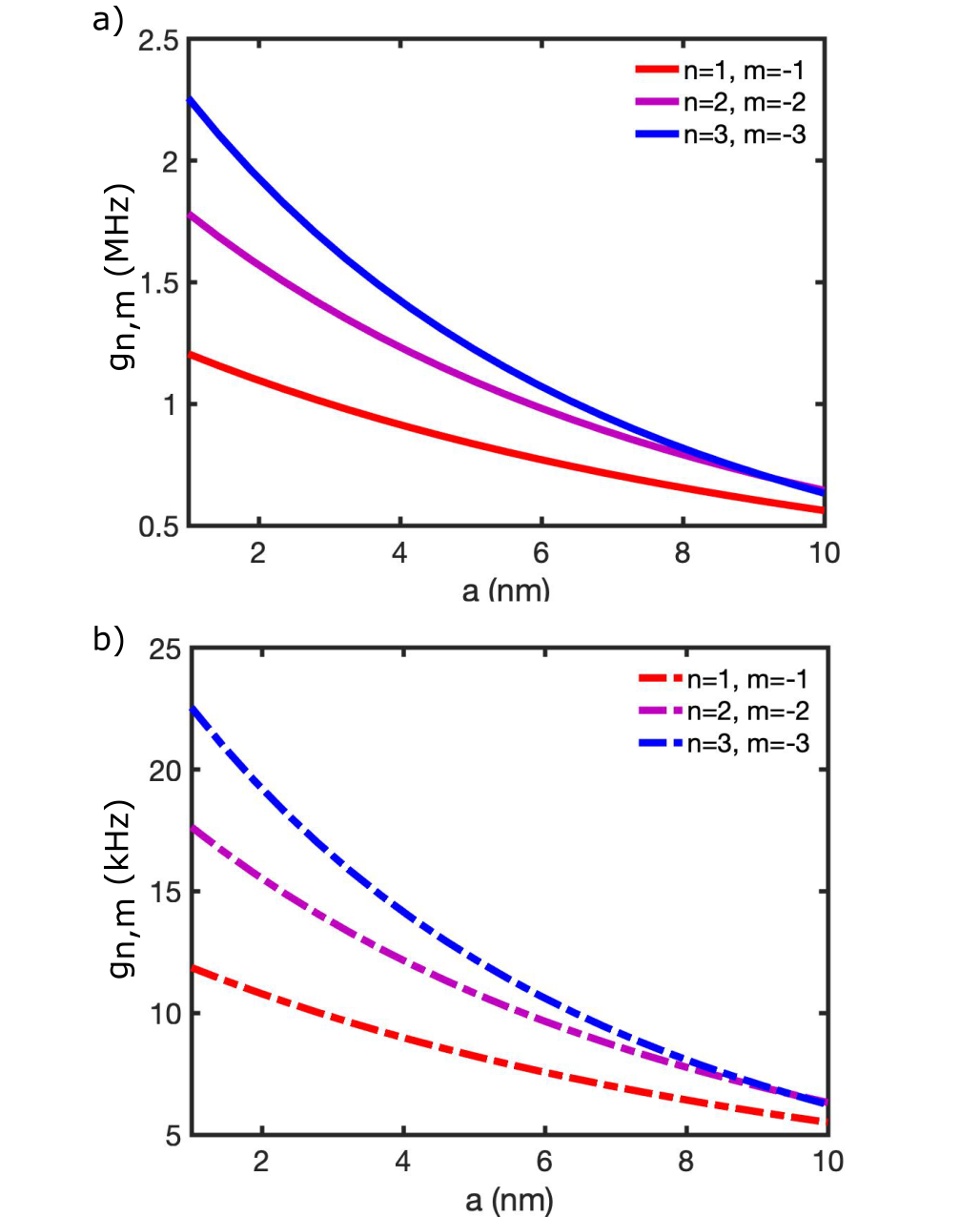}
\caption{
Coupling rate $g_{n,m}$ between the dipolar, quadrupolar, and octupolar modes in red, purple, and blue, respectively, for the \textbf{(a)} allowed $\psi_{e^+}(\mathbf{r})|\uparrow\rangle\rightarrow \psi_{e^+}(\mathbf{r})|\downarrow\rangle$ and \textbf{(b}) forbidden transition $\psi_{e^+}(\mathbf{r})|\uparrow\rangle\rightarrow \psi_{e^-}(\mathbf{r})|\downarrow\rangle$ of SiV$^-$ for varying distance $a$ from the surface of the magnetic nanoparticle at $\theta=\pi/2$ and $\phi=0$. Note that the y-axes are in different units.
}
\label{fig:coupling}
\end{figure}

We couple the SiV$^-$ spin emitter to equatorial multipolar modes $n=-m$ of the magnonic cavity realized by a nanomagnetic sphere to demonstrate non-zero, distance-dependent coupling $g$ between the forbidden spin transition and magnon modes in Fig. \ref{fig:coupling}(b). As a comparison, we also calculate the coupling rate $g$ to the magnetic dipole-allowed transition between $\psi_{e^+}(\mathbf{r})|\uparrow\rangle$ and $\psi_{e^+}(\mathbf{r})|\downarrow\rangle$ in Fig. \ref{fig:coupling}(a). For the allowed transitions coupled to the dipolar, quadrupolar, and octupolar modes in Fig. \ref{fig:emitter}(a), we find $g \sim 1$ MHz, where the ordering of coupling rates with respect to indices $n$ and $m$ mirror the trends in $J(\omega)$ in Fig. \ref{fig:cavity}(b). Given that the coupling rate $g$ is much larger than the damping $\Gamma$, this transition lies within the strong-coupling regime when the magnon mode is resonantly tuned to the transition, resulting in, for instance, Jaynes-Cummings-like coherent energy transfer between the singly-excited magnon mode and spin emitter in accordance with calculations in Ref. \cite{Neuman2020Nanomagnonics}. In Fig. \ref{fig:coupling}(b) plotting $g$ as a function of sphere-emitter distance $a$ for the forbidden transition, the ordering of coupling rates with respect to indices $n$ and $m$ identically mirror the trends in Fig. \ref{fig:coupling}(a). However, the magnitudes of $g\sim 10$ kHz for the forbidden transition are two orders of magnitude lower than $g$ of the allowed transition and four orders of magnitude lower than the damping rate $\Gamma$, placing this transition within the weak-coupling regime and enabling incoherent, accelerated energy transfer from the singly excited magnon mode to the spin emitter's $|1\rangle$ from the $|0\rangle$ state, or vice versa. As the cavity-enhanced decay rate follows $g^2/\Gamma$, we expect decay rates $\sim$1 kHz for the forbidden transition, as compared to no decay whatsoever in a vacuum according to conventional selection rules. Lower-loss magnetic materials, such as V[TCNE]$_x$ with magnon damping rate two orders of magnitude lower than YIG's \cite{Candido2020}, could be used to boost the nanomagnonic cavity-enhanced decay rate of the spin emitter and realize the strong-coupling regime where $g\gg \Gamma$. By comparison, in spin emitters realized by defects in solid-state materials \cite{Narang2019}, phonon fields with frequencies also in the microwave range can be used to drive spin transitions \cite{Bennett2013} with Rabi frequencies on the order of 1-50 MHz \cite{Meesala2018, Maity}, but these interactions can require careful control and engineering of phononic cavities.

We justify the assumption that the allowed and forbidden spin transition frequencies can be tuned in resonance with the magnon modes of interest. The spin transition energy $E_{\rm s}$ in the presence of a total magnetic field $B_z$ along the direction of the spin eigenbasis is \cite{Hepp2014,lemonde2018phononnetworks}
\begin{align}
    E_\mathrm{s} = (f\gamma_L+\gamma_S)H_z/\mu_0,
\end{align}
where the orbital gyromagnetic ratio $\gamma_L=\mu_B$, the spin gyromagnetic ratio $\gamma_S=2\mu_B$, and the coefficient of the orbital Zeeman quenching effect $f\approx -0.1$. The total magnetic field $H_z=|\bm{H}_\mathrm{e}|+|\bm{H}_\mathrm{sphere}|=|\bm{H}_\mathrm{e}^\mathrm{em}|$, where all magnetic fields point along the $z$-direction, the external magnetic field applied to the spin emitter SiV$^-$ $\mu_0\bm{H}_\mathrm{e}^\mathrm{em}$ can be chosen such that the spin transition frequency is resonant with the magnon frequencies plotted in Fig. \ref{fig:cavity}, and  $\mu_0 \bm{H}_\mathrm{sphere}$ is equivalent to the magnetic field of a magnetic point dipole at the center of the sphere with magnetic dipole moment $\bm{M}_s V$. 
Due to the magnetic field of the sphere acting on the spin emitter in addition to the externally applied magnetic field, there is an frequency gap between the magnon eigenfrequencies and the frequency of the forbidden transition of the spin emitter of $\sim 1$ GHz. Therefore, for this particular sphere-emitter setup, we require a difference of $\mu_0 \bm{H}_\mathrm{e}^\mathrm{em}\sim 35$ mT in externally applied magnetic field between the sphere and the spin emitter separated by 10 nm, feasible in recent demonstrations of microcoils fabricated on diamond chips with electrically tuneable magnetic field gradients of up to $\sim 10$ mT/nm \cite{Zhang2017, Jakobi2017, Bodenstedt2018}.

\section{Conclusion and outlook}

The present study provides the theoretical basis for breaking selection rules in spin emitters with magnetic nanoparticles. Specifically, we show that magnetic nanoparticles serving as nanomagnonic cavities to concentrate microwave magnetic fields generate fields with spatial gradients with length scales on the order of the spatial delocalization of spin emitters. As an example, we couple magnon modes of a YIG nanosphere to the SiV$^-$ defect center and show coupling on the order of tens of kHz between several equatorial, multipolar modes of the nanosphere and the forbidden transition between the canonical spin-based qubit states. Natural extensions of the present work include exploration of the coupling between planar interfaces and spatially delocalized emitters, which can support larger magnetic field gradients, analogous to the large electric near-fields supported by plasmonic surfaces.

\section*{Acknowledgements}
We acknowledge fruitful discussions with Christopher J. Ciccarino, Isaac Harris, and Michael Haas. This material is based upon work supported by the U.S. Department of Energy, Office of Science, National Quantum Information Science Research Centers. This work was also partially supported by the Department of Energy `Photonics at Thermodynamic Limits' Energy Frontier Research Center under grant DE-SC0019140. D.S.W. is supported by the Army Research Office MURI (Ab-Initio Solid-State Quantum Materials) grant number W911NF-18-1-0431 and the National Science Foundation Graduate Research Fellowship. P.N. is a Moore Inventor Fellow through Grant GBMF8048 from the Gordon and Betty Moore Foundation. This research used resources of the National Energy Research Scientific Computing Center, a DOE Office of Science User Facility supported by the Office of Science of the U.S. Department of Energy under Contract No.DE-AC02-05CH11231.

\appendix

\section{SiV$^-$ orbitals} \label{app:orbitalextraction}
To extract spin-orbit-coupled orbitals of SiV$^-$, we use the periodic density functional theory (DFT) code Quantum ESPRESSO \cite{Giannozzi2009, Espresso2017, Phys2020} with a geometry-optimized 511-atom supercell of diamond with a SiV$^-$ defect, the Perdew-Burke-Ernzerhof (PBE) exchange-correlation functional \cite{Perdew1996}, non-collinear spin-orbit coupling, kinetic energy cutoff of 80 Rydberg, and energy and force convergence thresholds of $10^{-8}$ a.u. and $10^{-5}$ a.u., respectively. The resulting four Kohn-Sham spin orbitals within the band gap--$\Psi_A(\mathbf{r}), \Psi_B(\mathbf{r}), \Psi_C(\mathbf{r}), \Psi_D(\mathbf{r})$--are split by spin-orbit coupling into two degenerate pairs, such that $E_A=E_B < E_C=E_D$. We can extract the spatial wavefunctions $\psi_{e^+}(\mathbf{r}), \psi_{e^-}(\mathbf{r})$ from either pair, but for specificity, we focus on the two Kohn-Sham orbitals with lower energies and note that we can equivalently apply this approach to the pair with higher energies. The lower energy pair is by default output as
\begin{align}
    \Psi_A(\mathbf{r})= a_z(\mathbf{r}) |\uparrow_z\rangle + b_z(\mathbf{r}) |\downarrow_z\rangle, \\
    \Psi_B(\mathbf{r}) = c_z(\mathbf{r}) |\uparrow_z\rangle + d_z(\mathbf{r}) |\downarrow_z \rangle,
\end{align}
where $a_z(\mathbf{r})$ to $d_z(\mathbf{r})$ are spatially-varying coefficients for the spin basis along the $z$-direction $|\uparrow_z\rangle$ and \mbox{$|\downarrow_z\rangle$}. However, we seek orbitals spin-polarized along the [111] direction, so we rotate the spinors by applying the spinor rotation matrix
\begin{equation}
    \bm{R} = \exp\left(\frac{-{\rm i}\bm{\sigma}\cdot \bm{n} \Delta \phi}{2}\right),
\end{equation}
where $\bm{\sigma}$ is the Pauli vector, $\bm{n}$ is the axis of rotation, and $\Delta\phi$ is the angle of rotation. We can therefore re-write $ \Psi_A(\mathbf{r})$ and $\Psi_B(\mathbf{r})$ as
\begin{align}
     \Psi_A(\mathbf{r}) = a(\mathbf{r})|\uparrow\rangle + b(\mathbf{r})|\downarrow \rangle, \\
     \Psi_B(\mathbf{r}) = c(\mathbf{r})|\uparrow\rangle + d(\mathbf{r})|\downarrow\rangle,
\end{align}
where $a(\mathbf{r})$ to $d(\mathbf{r})$ are spatially-varying coefficients for the spin basis along the $[111]$ direction $|\uparrow\rangle$ and $|\downarrow\rangle$. Due to spin-orbit coupling, the spatial coefficients $a$ and $c$ (or $b$ and $d$) corresponding to the $|\uparrow\rangle$ ($|\downarrow\rangle$) basis can be separated into the desired spatial wavefunctions $\psi_{e^+}(\mathbf{r})$ ($\psi_{e^-}(\mathbf{r})$) defined to a phase:
\begin{align}
    \Psi_A(\mathbf{r}) = \alpha \psi_{e^+}(\mathbf{r}) |\uparrow\rangle + \beta \psi_{e^-}(\mathbf{r}) | \downarrow\rangle, \\
    \Psi_B(\mathbf{r}) = \gamma \psi_{e^+}(\mathbf{r}) |\uparrow\rangle + \delta \psi_{e^-}(\mathbf{r}) |\downarrow\rangle,
\end{align}
where $\alpha$ to $\delta$ are complex numbers fulfilling $\alpha\gamma^*+\beta\delta^*=0$ and $|\alpha|^2+|\beta|^2=|\gamma|^2+|\delta|^2=1$.
\newcommand{\noopsort}[1]{} \newcommand{\printfirst}[2]{#1}
  \newcommand{\singleletter}[1]{#1} \newcommand{\switchargs}[2]{#2#1}

\appendix

\end{document}